\documentstyle[12pt]{article}

\begin{document}

\hoffset = -1.5truecm
\voffset = -4truecm

\title{\Large \bf
Chaotic Repellers in Antiferromagnetic Ising Model
}

\author{
N. S. Ananikian,$^{(a)}$ S. K. Dallakian, N. Sh. Izmailian, \\ and \\
K. A. Oganessyan \\
\normalsize Department of Theoretical Physics, Yerevan Physics Institute,
Alikhanian Br.2,\\ \normalsize 375036 Yerevan, Armenia }

\newpage

\maketitle

\begin{abstract}
For the first time we present the consideration of the antiferromagnetic Ising
model in case of fully developed chaos and obtain the exact connection between
this model and chaotic repellers. We describe the chaotic properties of this
statistical mechanical system via the invariants characterizing a fractal set
and show that in chaotic region it displays phase transition at {\it positive}
"temperature" $ \beta_c = 0.89 $. We obtain the density of the invariant
measure on the chaotic repeller.
\end{abstract}

\vspace*{1.0cm}


\vspace*{6.0cm}

$^{(a)}$ Electronic address: ananikian@vxc.yerphi.am

\newpage

It is now widely recognized that fractal objects play a fundamental role in
many branches of physics \cite{A1,A2,A3}. Much progress has been made
in finding an appropriate description, in terms of "chaotic repellers" or
"termodynamic formalism" for these structures (see ref. \cite{A4} and
references therein). The main goal of this Letter is the investigation of the
real statistical mechanical system in case of fully developed chaos.

Recently the three-site antiferromagnetic Ising spin model (TSAI) on Husimi
tree have been investigated. First it was shown that this approach yields good
approximation for the ferromagnetic phase diagrams \cite{A5}, which closely
match the exact results obtained on a Kagome lattice \cite{A6}. Second, the
quantitative picture of full doubling bifurcations diagram including
chaos for the magnetization of this system with finite coordination number
was obtained \cite{A7}. Note also, that in the
anisotropic-next-nearest-neighbor
Ising model on a Cayley tree in the infinite coordination limit the existence
of
chaotic phases associated with strange attractors have been obtained \cite{A8}.

With "termodynamic formalism" we investigate this statistical system in
chaotic region and describe the chaotic properties of it via the invariants
characterizing a fractal set (e.g. a strange attractor). In the present work
we obtain the strong connection between TSAI system in case of full chaos and
chaotic repellers. In particular we obtain the density of the invariant measure
on the chaotic repeller and focused on whether the termodynamical quantities
have phase transition \cite{A9}. It is, in general, hard to determine such
behavior unambiguously by numerical methods if one dose not have further
arguments or exact solutions (as in case of $x \to 4x(1-x)$) to depend on.
However, there is a possibility to remedies these deficiencies if consider the
characteristic Lyapunov exponents as an order parameter, which will differ in
two phases. Here we manage to calculate this order parameter as a function of
the "temperature" and clearly show that the phase transition occurs at
{\it positive} "temperature", whereas in the fractal sets, generated by a class
of maps close to $x \to 4x(1-x)$ the phase transition occurs at negative
"temperatures".

 The pure Husimi tree \cite{A10} is characterized the $\gamma$-the number of
triangles, which go out of each site. The 0th-generation is a single central
triangle.

 The TSAI model in the magnetic field defined by the Hamiltonian
$$
H=-J_3^{'}\sum_{\triangle}{\sigma}_i{\sigma}_j{\sigma}_k-h^{'}\sum_i
{\sigma}_i ,
$$
where ${\sigma}_i$ takes values $\pm 1$, the first sum goes over all
triangular faces of the Husimi tree and the second over all sites. Besides we
denote $J_3=\beta J_3^{'} ,\ h=\beta h^{'} ,\ \beta=1/kT $,
where h--external magnetic field, T--temperature of the system and $J_3<0$
corresponds to the antiferromagnetic case (in all our numerical calculations
we will put $J_3 = -1$).

The advantage of the Husimi tree introduced is that for the models
formulated on it, exact recursion relation can be derived \cite{A7,A11}:
\begin{equation}
\label{R1}
x_n=f(x_{n-1}),\qquad f(x)=\frac{z{\mu}^2x^{2(\gamma-1)}+2\mu x^{\gamma-1}+z}
{{\mu}^2 x^{2(\gamma-1)}+2z\mu x^{\gamma-1}+1} ,
\end{equation}
where $z=e^{2J_3}, \quad \mu=e^{2h}\ and\ 0\le x_n \le1 $. The function $f(x) $
is unimodal: continuous, continuously differentiable and has one maximum at
$x^{*}= {\mu}^{ 1/{(1-\gamma) }} $ in $ [0,1] $. Note that $f(x^{*}) = 1$ for
any $\gamma$, $h$ and $T$. This function is nonhyperbolic (hyperbolicity for
$1D$ maps means that $1 < f' < \infty $ in all points) and maps the interval
$[0,1]$ onto $[z,1]$.

The $x_n$ has no direct physical meaning but through it one can express the
magnetization of the central base site:
$$
m=\langle \sigma_0 \rangle =\frac{e^hx_n^{\gamma}-1}{e^hx_n^{\gamma}+1} ,
$$
and other thermodynamic parameters, since we can say that the $x_n$
determine the states of the system.

How to link the TSAI system in case of fully developed chaos with strange
invariant objects (Contor sets) of the phase space, called chaotic repellers
or semiattractors? This is the question we shall focus on.
For this we need a natural partition and that is provided by the cylinders
(see ref. \cite{A4}).

The recursion relation given by eq.(\ref{R1}) when $\gamma = 4$ will have the
following form:
\begin{equation}
\label{R2}
x_n = f(x_{n-1}), \quad f(x)=\frac{z{\mu}^2x^6+2\mu x^3+z}{{\mu}^2 x^6
+2z\mu x^3+1} .
\end{equation}
In the case when equation $f(x)-x =0$ has only one solution there are no points
of sequence $x_n$ which can be escaped and therefore no cylinders rise. It
means that there is no possibility to study the chaotic region of TSAI system
in terms of "termodynamic formalism". However in some special cases, when
equation $f(x) - x = 0$ have two or three solutions there is a possibility to
avoid this problem.

Let us consider the more simple case when equation $f(x)-x=0$ have two
solutions: $x_0$ and $x_1$. So the point $x_0$ must satisfy to condition:
$f'(x_0) = 1$ .

On the other hand we demand that the values of the function eq.(\ref{R1})
coincide at points $x=1$ and $x=x_0$. From these conditions after some
calculations we obtain:
\begin{equation}
\label{R3}
x_0 = {\left [ \frac{3+ \sqrt {10} - \sqrt {15+ 6\sqrt {10}}}{2} \right ]}^2,
\end{equation}
and corresponding values of $\mu$ and $z$ are:
\begin{equation}
\label{R4}
\mu = \mu_0 = 215.5128, \quad z = z_0 = 0.01854.
\end{equation}
The second solution of equation $f(x)-x=0$ is:
\begin{equation}
\label{R5}
x_1=0.3129, \quad |f'(x_1)| = 2.68 .
\end{equation}
The plot of function with these values of $\mu_0$ and $z_0$ is shown in Fig.1.
One can see from Fig.1 that in this case sequence $x_n$ maps the interval
$[x_0, 1] $ onto $[x_0, 1] $ and the points which come to $x^{*} = \sqrt{x_0}$
are escaped.

As known the analytical criterion for local unstability is:
$$ | \frac{d}{dx_1} f(x_1) | > 1  . $$
One can see from eq.(\ref{R5}) that the mentioned condition of unstability
is fulfilled.

On the other hand following to ref. \cite{A7} for $\gamma =4$ and $z_0 =
0.01854$ we calculate the Feingenbaum exponent $\delta$ ($\delta = 4.669 $) and
also $\mu_{\infty} = 3.18 $ from which the chaotic behavior ensues and
sequence essentially never repeats itself. It shows that the value of $\mu_0$
of eq.(\ref{R4}) corresponds to chaotic region.

Thus we obtain the exact connection between the Ising spin model with multisite
antiferromagnetic interaction on Husimi tree in external magnetic field
(when $\gamma =4$) in case of "fully developed chaos" and chaotic repellers.

It is interesting to note, that if one lets $\gamma=3$ in eq.(\ref{R1})
rather then $\gamma=4$, the above situation changes dramatically. In this
case equation $f(x)-x = 0$ have only one solution in chaotic region. It means,
that there is no possibility to study the TSAI system in case of "crisis" in
terms of chaotic repellers.

For a crisis map given by eq.(\ref{R2}) and eq.(\ref{R4}) we want to describe
the scaling properties of the attracting set, which in this case is the entire
interval $I$: $ [x_0, 1] $. For this we need a natural partition and that is
provided by the cylinders (we follow here refs. \cite{A8, A12}). For an index
$n$, $I$ is partitioned into $2^n$ intervals or cylinders, these being the
segments where the sets of points having identical symbolic-dynamics sequences
of length $n$ with respect to the maximum point $x^{*} = \sqrt x_0$. The
inverse function $h = f^{-1}$ has two branches and the $n$-
cylinders are all the nth-order preimages of $I$. The length of cylinder is
denoted by $l_{\epsilon_{1}, \epsilon_{2}, \dots \epsilon_{n}}$ , where
$\epsilon \in \{ -1,1 \} $.

The partition function $Z(\beta)$ is defined as
\begin{equation}
\label{R6}
Z_n(\beta) = \sum_{\epsilon_{1}, \epsilon_{2}, \dots \epsilon_{n} }
e^{-\beta ln {l_{\epsilon_{1}, \epsilon_{2}, \dots \epsilon_{n}}}},
\end{equation}
where $\beta \in (-\infty, \infty)$ is a free parameter, the inverse
"temperature". In the limit $n \to \infty$ the sum behaves as
\begin{equation}
\label{R7}
Z_n(\beta) = e^{-nF({\beta})},
\end{equation}
which defines the free energy, $F(\beta)$ (the topological pressure
\cite{A13}).
The entropy $S(\lambda)$ is the Legendre transform $S(\lambda) = -F(\beta)
+{\lambda}{\beta}$,
where the relation between $\lambda$ and $\beta$ is found from $\lambda =
F'(\beta)$, $ \beta(\lambda) = S'(\lambda)$,
and these have the following meaning: in the limit $n \to \infty$,
$e^{nS(\lambda)} $ is the number of cylinders with length $l = e^{-n\lambda}$
or, equivalently, with Lyapunov exponent $\lambda$. The dimension of the set
of points in $I$ having Lyapunov exponent $\lambda$ is $S(\lambda) / \lambda$.

With using the eqs.(\ref{R6}), (\ref{R7}) we can numerically calculate
the free energy, which is shown in Fig.2. One can see from Fig.2 that the
free energy have a singularity around $\beta = 1$ and shows the existence
of the phase transition of first order in this region of $\beta$.

How can the transition be determined accurately?

Let us consider the characteristic Lyapunov exponent as an order parameter
which will differ in two phases. Fig.3 shows this order parameter at positive
temperatures calculated for different sizes of the system, corresponding to
$n = 9, 11, 13$. The curves converge towards the line and result in a first
order transition at positive $\beta_{c} = 0.89$, whereas in the fractal sets,
generated by a class of maps close to $x \to 4x(1-x)$ the phase transitions of
first order occur at negative $\beta_{c}$ \cite{A12}.

To consider the above obtained results in terms of entropy function
$S(\lambda)$, let us first discuss the general appearance of the entropy
function. First of all it should be positive on some interval $[\lambda_{min},
\lambda_{max} ] $. The value $\lambda = ln2$ must belong to that interval,
which
follows from the fact that the sum of the lengths of all cylinders on a given
level is 1. Secondly it is often found that the values of $\lambda_{min}$ and
$\lambda_{max}$ are given by the logarithms of the slopes at the origin
\cite{A8}. In our case $\lambda_{min} = 0$ ($f'(x_0) = 1 $) and
$\lambda_{max} = ln2.68$ ($|f'(x_1)| = 2.68$). Consequently the slope at the
fixed point away from $x_0$ ($x_1$) is larger then the slope at the origin
$x_0$.  It also indicates the quite different behavior of TSAI system then in
the cases of a class of maps close to $x \to 4x(1-x)$, where the phase
transition exists \cite{A12}.

The precise form of the entropy function is, as mentioned in the introduction,
not easy to assess with great accuracy. The existence of a first order
phase transition implies that there should be a straight line segment in
$S(\lambda)$. Combining the value of $\beta_{c} = 0.89$ with the point
corresponding to this $\beta_{c}$: $\lambda = 0.369, S(\lambda) = 0.414 $ we
find the straight line segment in $S(\lambda)$ when $n=12$. Of course,
with finite-size data, it is impossible determine the full straight line
segment in $S(\lambda)$ and with increasing the $n$, the straight line
segment will increase.

Finally let us consider the recursion \cite{A14}:
\begin{equation}
\label{R8}
Q_{n+1}^{(\beta)}(x') = e^{\beta F(\beta) } \sum_{x \in f^{-1}(x')}
\frac{Q_{n}^{(\beta)}(x)}{{|f'(x)|}^{\beta}} .
\end{equation}

In cases with a chaotic attractor, eq.(\ref{R8}) for $\beta = 1$
(It has been claimed \cite{A15} that for $1D$ maps with chaotic attractors
$F(1) = 0$) is the well known Frobenius-Perron equation \cite{A16}:
\begin{equation}
\label{R9}
Q_{n+1}(x') = \sum_{x \in f^{-1}(x')} \frac{Q_n(x)}{|f'(x)|} ,
\end{equation}
where $Q(x) = \lim \limits_{n \to \infty} Q_n(x)$ is the invariant density
on chaotic repeller or chaotic attractor. The iterative solution of
eq.(\ref{R9}) converges very fast ($n=5$) and we determine the stationary
solution of it, which is shown in Fig.4.

In summary we have obtained the exact connection between the Ising spin model
with multisite antiferromagnetic interaction on Husimi tree in external
magnetic
field (when $\gamma =4$) in case of "fully developed chaos" and chaotic
repellers. This remarkable result gives possibility to investigate the chaotic
properties of this statistical mechanical system via the invariants
characterizing a fractal set (e.g. a strange attractor). In particular we
numerically solved Frobenius-Perron recursion equation and obtained the density
of the invariant measure on the repeller. It is shown that this system in
chaotic region displays a first order phase transition at {\it positive}
"temperature" $ \beta_c = 0.89 $, whereas in the fractal sets, generated by a
class of maps close to $x \to 4x(1-x)$ the phase transitions of first order
occur at negative "temperatures". It is, in general, hard to determine such
behavior unambiguously by numerical methods if one dose not have further
arguments or exact solutions to depend on. However, there is a possibility to
remedies these deficiencies if consider the characteristic Lyapunov exponents
as an order parameter, which will differ in two phases. In this Letter we
manage
to calculate this order parameter as a function of the "temperature" and
clearly
show where the phase transition of first order will occur, which describe
transitions in the distribution of the characteristic Lyapunov exponents. This
phase transition in terms of free energy and entropy function is also analysed.

The investigation of chaotic statistical physical system has opened new
challenges for theories of stochastic processes, especially in the direction
of stochasticity of vacuum in QCD. Note that the $Z(3)$ gauge model with a
double plaquette representation of the action on the generalized Bethe lattice
is dual to the spin-one Blume-Emery-Griffiths model \cite{A17,A18}. This allows
us to suppose that TSAI on Husimi tree when $\gamma =4$ can be connected via
duality with multiplaquette representation of the gauge theory. Thus we hope
that the present work will stimulate the further investigation of stochastic
processes.

The authors are grateful to A. Akheyan, R. Flume and S. Ruffo for useful
discussions. One of the authors (N.A.) should like to acknowledge Professor
Abdus Salam, the International Atomic Energy Agency and UNESCO for
hospitality at the International Centre of Theoretical Physics. He would like
also to acknowledge to R. Markarian, Ya. Sinai for their interest in this work
and helpful discussions on "Workshop on Dynamical Systems" (I.C.T.P., Trieste,
22 May - 2 June 1995).

This  work  was partly  supported by ISF Supplementary Grant-SDU000 and
the Grant-211-5291 YPI of the German Bundesministerium  fur  Forshung
and Technologie.

\newpage

\begin{center}
Figure Captions
\end{center}

Fig.1. The function eq.(\ref{R2}) for values of $\mu_0 $ and $z_0$ of
eq.(\ref{R4}).

Fig.2. The free energy $F(\beta)$, which have a singularity around
$\beta = 1$ and shows the existence of the phase transition of first
order in this region of $\beta$.

Fig.3. The order parameter $\lambda(\beta)$ calculated for different sizes
of the system, corresponding to $n = 9, 11, 13$. The curves converge towards
the line and result in a first order transition at $\beta_{c} = 0.89$.

Fig.4. Invariant density on chaotic repeller.

\end{document}